\documentclass[a4paper,fleqn,usenatbib]{mnras}
\usepackage{newtxtext,newtxmath}

\usepackage[T1]{fontenc}
\usepackage{ae,aecompl}

\usepackage{cmap}
\usepackage{graphicx}	
\usepackage{amsmath, bm}	
\usepackage{amssymb}	

\usepackage{hyperref}

\makeatletter
\newcommand{\customlabel}[2]{%
   \protected@write \@auxout {}{\string \newlabel {#1}{{#2}{\thepage}{#2}{#1}{}} }%
   \hypertarget{#1}{#2}
}
\makeatother

\title[Whistler instabilities in space plasmas]{Whistler instabilities from the interplay of electron anisotropies
in space plasmas: A quasilinear approach}
\author[S.M.Shaaban]{
S. M. Shaaban,$^{1,2}$\thanks{E-mail: s.m.shaaban88@gmail.com} and M. Lazar,$^{1,3}$ 
\\
 $^{1}$Centre for Mathematical Plasma Astrophysics, KU Leuven, Celestijnenlaan 200B, B-3001 Leuven, Belgium.\\
$^{2}$Theoretical Physics Research Group, Physics Department, Faculty of Science, Mansoura University, 35516, Mansoura, Egypt.\\
$^{3}$Institut f\"ur Theoretische Physik, Lehrstuhl IV: Weltraum- und Astrophysik, Ruhr-Universit\"at Bochum, D-44780 Bochum, Germany. }
\date{Accepted XXX. Received YYY; in original form ZZZ}
\pubyear{2019}

\begin{document}
\label{firstpage}
\pagerange{\pageref{firstpage}--\pageref{lastpage}}
\maketitle

\begin{abstract}
Recent statistical studies of observational data unveil relevant correlations between 
whistler fluctuations and the anisotropic electron populations present in space plasmas, e.g., 
solar wind and planetary magnetospheres. Locally, whistlers can be excited by two sources of 
free energy associated with anisotropic electrons, i.e., temperature anisotropies and beaming 
populations carrying the heat flux. However, these two sources of free energy and the resulting instabilities 
are usually studied independently preventing a realistic interpretation of their interplay. This paper 
presents the results of a parametric quasilinear study of the whistler instability cumulatively driven by 
two counter-drifting electron populations and their anisotropic temperatures. By 
comparison to individual regimes dominated either by beaming population or by temperature anisotropy, 
in a transitory regime the instability becomes highly conditioned by the effects of both these two 
sources of free energy. Cumulative effects stimulate the instability and enhance the resulting
fluctuations, which interact with electrons and stimulate their diffusion in velocity space, leading 
to a faster and deeper relaxation of the beaming velocity associated with a core heating in perpendicular 
direction and a thermalization of the beaming electrons. In particular, the relaxation of temperature 
anisotropy to quasi-stable states below the thresholds conditions predicted by linear theory may explain 
the observations showing the accumulation of these states near the isotropy and equipartition of energy.

\end{abstract}


\section{Introduction}\label{sec.1}

Two prominent sources of free energy are revealed by the velocity distributions of plasma particles
in space plasmas, i.e., temperature anisotropies and beaming (or strahl) populations \citep{Pilipp1987, 
Crooker2003, Stverak2008, Vinas2010, Tong2019a}. These anisotropies are at the origin of various 
instabilities enhancing the wave fluctuations and turbulence detected at kinetic scales \citep{Sahraoui2009, 
Bale2009, Alexandrova2013, Wilson2013}. In collision-poor plasmas from space, the 
kinetic wave fluctuations play major roles, interacting resonantly with plasma particles and 
triggering not only their relaxation \citep{Saito2007, Shaaban2016, Gershman2017, Lazar2018WI, Lopez2019, 
Shaaban2019HF} but also the wave energy dissipation at small scales \citep{Leamon1998, Saito2008, 
Parashar2009, Goldstein2015}. In the present paper we investigate the whistler fluctuations 
\citep{Kennel1966, Gary1975, Garybook, Shaaban2018a, Lazar2019}, also known as electromagnetic electron cyclotron 
modes \citep{Cuperman1981}, and focus on the instability of these modes, cumulatively driven by (counter-)beaming 
electron populations and their temperature anisotropy. In this case linear theory shows significant 
changes of the instability conditions and growth rates \citep{Lazar2008, Shaaban2018a}, motivating
the interest for an extended quasilinear (QL) study to characterize the long-term evolution of growing 
fluctuations which interact with electrons and contribute to their relaxation. 
The observed whistler fluctuations have not only a sufficiently wide frequency width but also 
low amplitudes, comparing to the background magnetic field $B_0$, i.e., $\delta B^2 << B_0^2$ 
\citep{Tong2019a, Tong2019b}, which give more credits to a QL approach to provide valid descriptions of
these fluctuations and their action back on anisotropic electrons. A QL approach may therefore help 
to understand the observations of whistler like fluctuations which are often associated with combinations 
of counter-beaming electron populations and their temperature anisotropies, see \cite{Tong2019a, Tong2019b}.

The most popular is probably the whistler instability driven by the electrons with anisotropic temperature 
$T_\perp > T_\parallel$ (where $\perp$ and $\parallel$ denote directions with respect to the magnetic 
field) \citep{Gary1996}, recent studies showing also implications in the solar wind conditions 
\citep{Stverak2008, Lazar2018a, Bercic2019, Shaaban2019WI}. Thus, the fact that instability thresholds 
shape the observed temperature anisotropy of the non-drifting electron populations is an indirect proof 
of the constraining role that this instability may play in space plasmas \citep{Stverak2008, Lazar2018a, 
Bercic2019, Shaaban2019WI}. QL studies and numerical simulations confirm indeed the relaxation of 
anisotropic electrons to the same quasi-stable states predicted by the linear thresholds \citep{Sarfraz2016, 
Yoon2017, Kim2017, Shaaban2019WI}. However, the highest number of solar wind data concentrate below these 
thresholds, at lower anisotropies, and are usually explained invoking a collisional relaxation 
\citep{Salem2003, Stverak2008}, although particle-particle collisions become less efficient in the 
solar wind with increasing the heliospheric distance (e.g., at 1~AU and beyond).

On the other hand, the beaming or strahl population carrying the electron heat flux in the 
solar wind \citep{Maksimovic2005, Pagel2007, Gurgiolo2012, Bercic2019} is often associated 
with the enhanced fluctuations self-generated (locally) by the so-called whistler heat flux 
(WHF) instability \citep{Gary1975, Gary1985, Shaaban2018a, Shaaban2018b, Tong2019a, Shaaban2019HF, 
Lopez2019}. In this case the instability conditions are markedly restrained, e.g., to low beaming 
velocities, and to a strahl population less dense but hotter than the main ele/ctron population 
\citep{Gary1985, Shaaban2018b}. That may explain difficulties encountered in describing the long-term 
evolution of the WHF instability \citep{Shaaban2019HF, Lopez2019}. However, recent reports from 
QL analysis \cite{Shaaban2019HF} and numerical simulations \citep{Lopez2019} have explained the 
saturation of WHF instability by two synchronous effects, namely, a minor relaxation of the 
relative drifts combined with a small temperature anisotropy ($T_{b, \parallel} \gtrsim T_{b,\perp}$) 
effectively induced to the beaming population. The observations confirm the existence of these whistler 
fluctuations showing also evidences of their suppression even for a moderate temperature 
anisotropy \citep{Tong2019a, Tong2019b}. These recent results suggest that WHF 
instability cannot efficiently scatter and isotropize the strahl electrons, and therefore it cannot 
regulate the electron heat flux in the solar wind \citep{Shaaban2019HF, Lopez2019, Kuzichev2019}. 
However, these results undermine long-established thoughts which invoke this instability to explain 
long series of observations showing the decrease of relative density and pitch-angle scattering of 
the electron stahl with heliospheric distance \citep{Maksimovic2005, Pagel2007, Gurgiolo2012, Bercic2019}, 
and also an electron heat-flux below a collisional level \citep{Bale2013}, not consistent with the 
conventional Spitzer-H{\"a}rm predictions~\citep{Spitzer1953}. 
The mechanisms involving the self-induced instabilities need therefore further 
exploration in order to understand their implications and explain the observations. Here we assume 
less idealized conditions, which combine beaming electron populations with intrinsic temperature 
anisotropies \citep{Vinas2010, Tong2019a} and may, thus, trigger new regimes of whistler 
instabilities \citep{Lazar2008, Shaaban2018a}. 

Such unstable states have been described in linear theory. For instance, if 
whistler instability is mainly driven by temperature anisotropy of electrons (e.g., $T_\perp > T_\parallel$), 
the growth rates are inhibited by (increasing) the relative drift between electron core and beam populations 
\citep{Lazar2008, Shaaban2018a}. On the other hand, the influence of temperature anisotropy on the 
WHF instability depends on the nature of that anisotropy. Thus, linear growth rates are stimulated when 
beaming population (subscript $b$) exhibit $T_{\perp b} > T_{\parallel b}$, but instability is inhibited 
by an opposite anisotropy $T_{\parallel b} > T_{\perp b}$ \citep{Shaaban2018a}. However, a linear approach 
cannot describe more complex effects deriving from the saturation of the instability and the effects of 
the enhanced whistler fluctuations back on electron velocity distributions.

Here we present a QL analysis able to characterize not only the linear growth, but the long-term evolution 
of the whistler-like instability resulting from the interplay of (counter-)beaming electron populations and 
their temperature anisotropy. Our results enable to identify the contribution of these instabilities to the
relaxation of electron distributions, quantifying the time variations of both these sources of free energy.
In section \ref{Sec:2} we briefly introduce the linear and QL theory of whistler instabilities for such 
complex conditions of plasma electrons. Numerical solutions for the WHF instability under the effects of 
temperature anisotropies of the beam and core populations are discussed in section~\ref{sec.3.1}. In 
section~\ref{sec.3.2} we analyze the complementary regime of whistler instability mainly driven by 
temperature anisotropies, but under the influence of small (counter-)drifts which may alter predictions 
made for the temperature anisotropy limits of nondrifting plasma populations. Section \ref{Sec.4} summarizes 
the results of the present study and discusses their importance, in particular, for a better understanding 
of the observed whistler-like fluctuations and their potential implications in the isotropization of electrons 
in the solar wind.

\section{Quasi-linear Approach}\label{Sec:2}

In a workframe fixed to protons the counter-moving electron populations, namely, the core (subscript "$c$") and the beam 
(subscript "$b$") are generically described by the velocity distribution
\begin{align}
f_e(v_\perp, v_\parallel)=\dfrac{n_c}{n_0} f_c\left(v_\perp, v_\parallel\right)+\dfrac{n_b}{n_0}  f_b\left(v_\perp, v_\parallel\right),   \label{1}
\end{align}
where $n_b/n_e=\delta$ and $n_b/n_e=1-\delta$ are relative number densities and $n_0\equiv n_e = n_c + n_b\approx n_p$ is the total density.
In order to investigate the cumulative effects of electron beams and temperature anisotropy, we assume the electron populations 
described by drifting bi-Maxwellian distribution functions 
\begin{equation}
f_{j}(v_\perp, v_\parallel)=\frac{1}{\pi^{3/2}\alpha_{\perp~j}^{2}\alpha_{\parallel ~j}}
\exp \left(-\frac{v_{\perp }^{2}}{\alpha_{\perp ~j}^{2}} -
\frac{\left(v_\parallel -U_j\right)^{2}}{\alpha_{\parallel ~j}^{2}}\right),  \label{e2}
\end{equation}
with thermal velocities $\alpha_{\perp, \parallel ~j}(t)=~\sqrt{2k_{B}T_{\perp,\parallel~j}/m_{j}}$ 
defined in terms of the kinetic temperature components, perpendicular ($T_\perp$) and parallel ($T_\parallel$) 
to the stationary magnetic field $\bm{B}_0$. Parallel drifting velocities $U_j$ are conditioned by $n_c U_c~+~n_b 
U_b=0$, in order to maintain a zero net current. 

For a collisionless and homogeneous electron-proton plasma, the whistler modes manifest instabilities with maximum 
growth rates for propagation parallel to the stationary magnetic field, i.e.\ $\bm{k}\times \bm{B}_0=0$. For the parallel
electromagnetic modes the instantaneous linear dispersion relation reads  \citep{Gary1985, Shaaban2018a}
\begin{align}
c^{2}k^{2} =\omega ^{2}+\sum_{j}\omega _{e}^{2}\left[\xi_j 
Z\left( \zeta _j\right)+\left(\frac{T_{\perp j}}{T_{\parallel j}}-1\right)\left\{ 1+\zeta_j Z\left( \zeta _j \right) \right\} \right],\label{e3}
\end{align}
where $k$ is the wave number, $\omega_{e }=(4\pi n_{0} e^{2}/m_{e})^{0.5}$ is the electron plasma frequency, $\omega\equiv\omega_r+i\gamma$ is the wave frequency (represents the complex solution of the dispersion relations), $c$ is the speed of light,  $T_{\perp j},/T_{\parallel j}\equiv A_j$ is the temperature anisotropy, $\xi_{j}=~\left(\omega -kU_j\right) k^{-1}\alpha_{\parallel~j}^{-1}$, and 
\begin{equation}
Z\left( \zeta _j\right) =\frac{1}{\sqrt{\pi}}\int_{-\infty
}^{\infty }\frac{e^{ -x^{2}}}{x-\zeta_j}dt,\ \
\Im \left( \zeta_j^{\pm }\right) >0  \label{e4}
\end{equation}
is the transcendental plasma dispersion function \citep{Fried1961} of argument
\begin{equation*}
\zeta_{j }=\frac{\omega-\Omega _e-kU_j}{k\alpha_{\parallel~j}}.
\end{equation*}

In a quasi-linear (QL) formalism, the general kinetic equation for the electron distributions 
in the diffusion approximation takes the following form \citep{Yoon2017}
\begin{align} \label{e6}
\frac{\partial f_j}{\partial t}&=\frac{i e^2}{4m_j^2 c^2~ v_\perp}\int_{-\infty}^{\infty} 
\frac{dk}{k}\left[ \left(\omega^\ast-k v_\parallel\right)\frac{\partial}{\partial v_\perp}+ 
k v_\perp\frac{\partial}{\partial v_\parallel}\right]\nonumber\\
&\times~\frac{ v_\perp \delta B^2(k, \omega)}{\omega-kv_\parallel-\Omega_j}\left[ 
\left(\omega-k v_\parallel\right)\frac{\partial}{\partial v_\perp}+ k v_\perp
\frac{\partial}{\partial v_\parallel}\right]f_j
\end{align}
with the spectral wave energy of the fluctuations $\delta B^2$ described by the wave kinetic 
equation
\begin{equation} \label{e8}
\frac{\partial~\delta B^2(k)}{\partial t}=2 \gamma_k \delta B^2(k),
\end{equation}
where $\gamma_k$ is the instantaneous growth rate of whistler instabilities derived from Eq.~\eqref{e3}.  
The time evolution of the macroscopic moments of the eVDFs such that the temperature components 
$T_{\perp, \parallel  j}$ of  beam (subscript "$j=b$") and core (subscript "$j=c$") and their drift 
velocities $U_j$ is then governed by the following QL kinetic equations
\begin{subequations}\label{e7}
\begin{align}
\frac{dT_{\perp j}}{dt}&=\frac{1}{2}\frac{\partial}{\partial t}\int d{\bf{v}} ~m_j v_{\perp}^2~f_j(v_\perp, v_\parallel)\\
\frac{dT_{\parallel~j}}{dt}&=\frac{\partial}{\partial t}\int d{\bf{v}}~m_j(v_{\parallel}-U_j)^2~f_j(v_\perp, v_\parallel)\\
\frac{d U_j}{d t}&=\frac{\partial}{\partial t}\int d{\bf{v}}~ v_{\parallel}~f_j(v_\perp, v_\parallel)
\end{align}
\end{subequations}
The QL approach used in the present study is based on these equations. For the sake of completeness, including 
full mathematical derivations of Eqs.~\ref{e7}, the interested reader may refer to previous studies by 
\cite{Moya2011}, \cite{Yoon2017}, \cite{Shaaban2019HF, Shaaban2019WI} and refs therein. 
%
\begin{figure*}
\includegraphics[scale=1.1, trim={2.75cm 11.8cm 2.3cm 2.55cm}, clip]{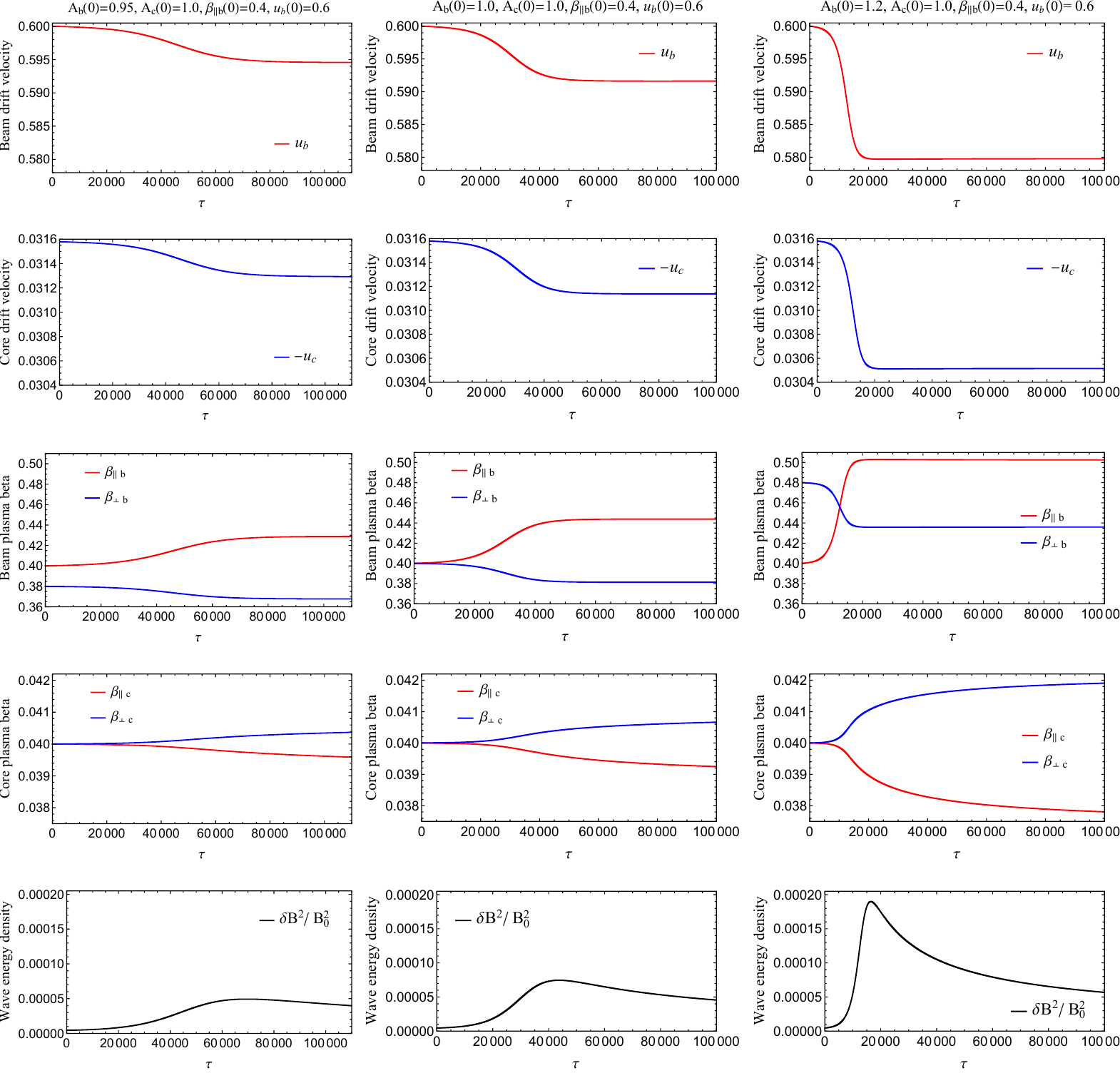}
\caption{Case~\ref{c1}: Effects of the initial beam temperature anisotropies $A_b(0)=0.95, 1, 1.2$ on the instability and saturation of the 
whistler wave energy $(\delta B^2/ B_0^2)$, and the relaxation of the plasma beta parameters ($\beta_{b,\parallel,\perp} 
\text{ and }\beta_{c,\parallel,\perp}$), and the drift velocities $\left(u_{b, c}\right)$.}\label{f1}
\end{figure*}
%

\section{Numerical solutions} \label{Sec:3}
%
In this section we solve the QL equations \ref{e7} and \ref{e8} numerically for four different sets of initial macroscopic plasma parameters (at $\tau=0$), namely, cases \ref{c1}, \ref{c2}, \ref{c3}, and \ref{c4}.
\begin{eqnarray}
&&\text{Case \customlabel{c1} {\color{blue} 1}}: A_{b}(0)=0.95, 1.0, 1.2, A_{c}(0)=1.0,  u_{b}(0)=0.6, \nonumber\\
&&\text{Case \customlabel{c2}{\color{blue} 2}}: A_{c}(0)=A_{b}(0)=0.95, 1.0, 1.2, u_{b}(0)=0.6, \nonumber\\
&&\text{Case \customlabel{c3}{\color{blue} 3}}: u_{b}(0)=0.0, 0.55, 0.7, A_{b}(0)=3, A_{c}(0)=1, \nonumber\\
&&\text{Case \customlabel{c4}{\color{blue} 4}}: u_{b}(0)=0.0, 0.55, 0.7, A_{b}(0)= A_{c}(0)=3,\nonumber
\end{eqnarray}
and $\delta=0.05, \beta_b(0)=0.4, T_b(0)=10~T_c(0), W(k)=5\times10^{-6}$, $u_{j}=\mu^{-0.5}U_j/v_A$, where 
$\mu=m_p/m_e=1836$ is the proton-electron mass ratio and  $v_A=2\times 10^{-4} c$ is the proton Alfv{\'e}n speed. 
QL analysis enables us to study temporal profiles of the enhanced wave fluctuations associated with the 
instability saturation, as well as their back reaction on macroscopic plasma 
parameters such as the beam and core plasma betas ($\beta_{\parallel b,c}$, $\beta_{\perp b,c}$), temperature 
anisotropies ($A_{b,c}$), and the corresponding drift velocities ($u_{b,c})$.

\subsection{Whistler heat-flux instability}\label{sec.3.1}
Here we focus on the unstable WHF solutions resulted as a cumulative effect of the relative drift velocities 
of the beam and core components and their temperature anisotropies. In cases \ref{c1} and \ref{c2} we adopt 
initial conditions favorable to WHF instability by assuming small temperature anisotropies for both the 
beam and core populations. 

\subsubsection{Case~\ref{c1} -- anisotropic beam electrons, $A_b(0)\neq 1$} 
%
\begin{figure*}
\centering
\includegraphics[scale=0.76, angle=-90, trim={6.5cm 4.cm 6.3cm 2.4cm}, clip]{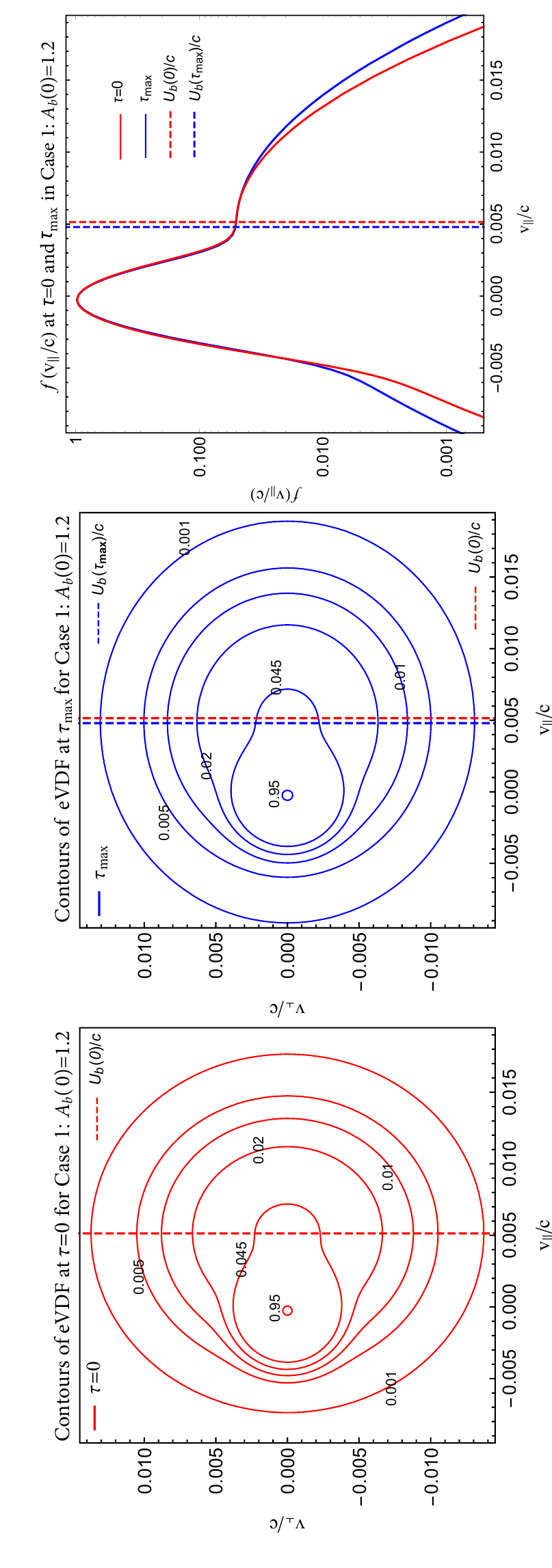}
\caption{Contours (left and middle panels) and parallel cuts (right) of the eVDF at initial 
$\tau=0$ (red) and final $\tau_m$ (blue) stages from the same QL run for case~\ref{c1} with $A_b(0)=1.2$ in 
Fig.~\ref{f1}.}\label{f2}
\end{figure*}

Initial plasma parameters are carefully selected in this case, adopting relatively small anisotropies 
$A_b(0)\equiv \beta_{\perp~b}(0)/\beta_{\parallel~b}(0) =0.95, 1.2$, and a small plasma beta 
$\beta_{\parallel~b}(0)=0.4$ for the beam, which guarantee the dominance of WHF instability and to 
avoid a major effect of the temperature anisotropy driven instabilities. Thus, Figure~\ref{f1} shows
temporal ($\tau=|\Omega_e|t$) evolutions of the parallel (red) and perpendicular (blue) plasma beta 
parameters for the the beam ($\beta_{\parallel, \perp~b}$) and for the core ($\beta_{\parallel, \perp~c}$), 
their drift velocities  ($u_{b,c}$), and the associated magnetic wave energy ($\delta B^2/B_0^2$) for 
three initial conditions given by different temperature anisotropies $A_b(0)=0.95$ (left), $1.0$ (middle), 
$1.2$ (right). For reference to previous studies, in the middle panels we consider the case of an initially 
isotropic beam ($A_b(0)=1.0$). The saturation of the WHF instability occurs from a concurrent effect of a 
relatively small relaxation of drift velocities with temperature anisotropies induced in the beam and core 
electrons \citep{Shaaban2019HF}. For $A_b(0)>1.0$ the instability develops faster 
and magnetic wave energy reaches a maximum level almost four times higher than that obtained for beams 
with $A_b(0) = 1$. The enhanced fluctuations determine in this case more pronounced effects on the electron 
populations, e.g., faster and deeper relaxation of the beaming velocity, and enhanced selective cooling or
heating of the core and beam populations. The relaxation of the beaming velocity remains however modest, 
but we may expect more stable distributions after saturation. Thus, for $A_b(0)=1.2$ the relaxation shows 
a quite interesting evolution, including a significant decrease of this anisotropy reaching an isotropic 
state $A_b =1$ before changing to opposite anisotropy $A_b(t_{m})<1.0$, later at $\tau = \tau_{m}$. 
Figure~\ref{f2} displays the normalized eVDFs corresponding to the QL run in case~\ref{c1} with $A_b(0)$=1.2 
(see Figure~\ref{f1}), as contours in ($v_\perp/c - v_\parallel/c$)$-$velocity space (left and middle) and 
their parallel cuts $f_e(v_\parallel/c)$ (right panel) at initial $\tau =0$ (red) and $\tau =\tau_m$ (blue) 
after saturation. Shown are the following contours $10^{-3},~5\times 10^{-3},~10^{-2},~0.02,~0.045$, and $0.95$. 
Indeed, Figure~\ref{f2} indicates only a minor relaxation of the beaming velocities, see the slightly visible 
difference between red ($\tau=0$) and blue ($\tau_m$) dashed lines, and shows the eVDF changing the anisotropy 
due to thermalization in parallel direction (blue solid line in the left panel). This may explain the induced 
temperature anisotropy of the beam component. As one can see in the middle panel, compared to initial state 
the eVDF becomes less asymmetric in perpendicular direction and therefore more stable against WHF instability.

%
%
\begin{figure*}
\includegraphics[scale=0.76, angle=-90, trim={8.6cm 3.6cm 8.cm 2.4cm}, clip]{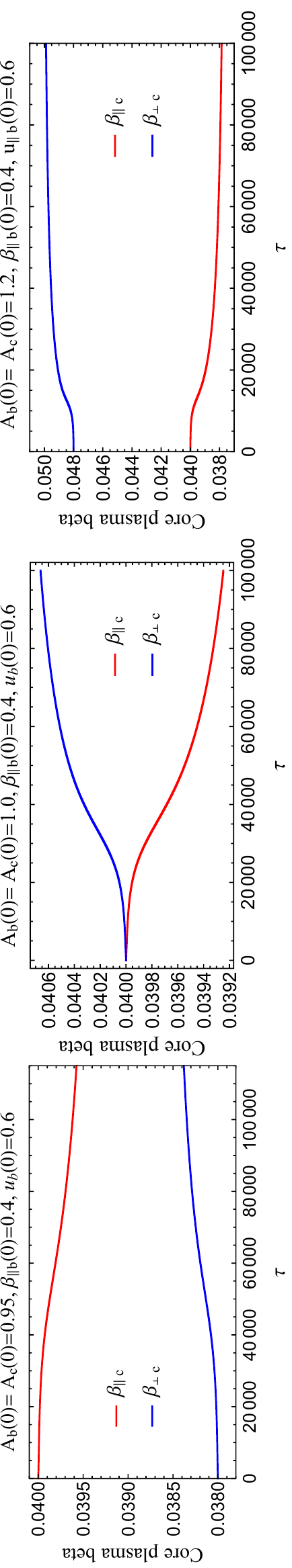}
\caption{Case~\ref{c2}: The same as in Figure~\ref{f1}, but with initially anisotropic core $A_c(0)=A_b(0)$.}\label{f3}
\end{figure*}

\begin{figure*}
\includegraphics[scale=1.1, trim={2.75cm 11.5cm 2.3cm 2.55cm}, clip]{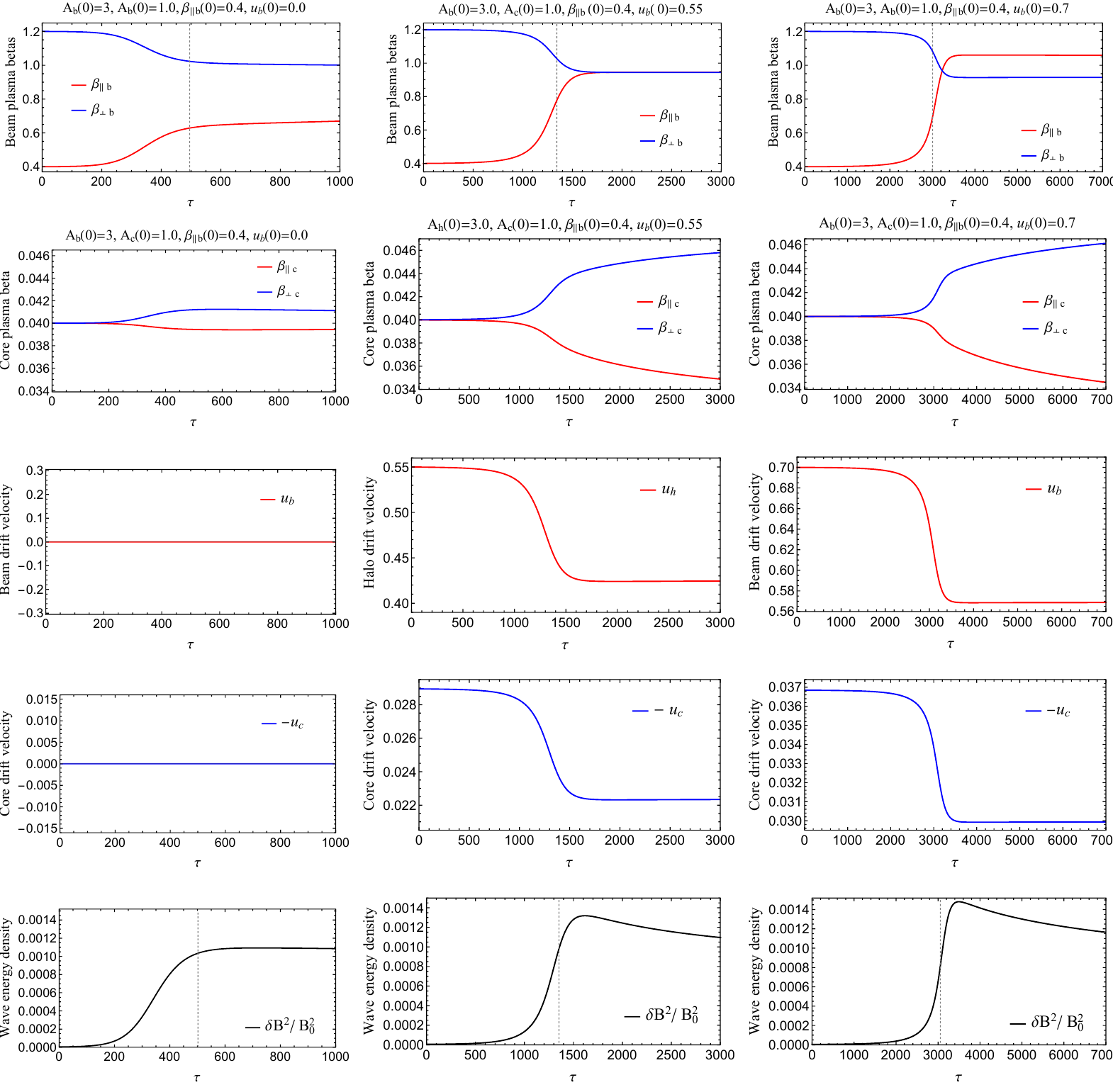}
\caption{Case~\ref{c3}: Effects of the initial beam drift velocity $u_h(0)=0.0, 0.55, 0.7$ on the whistler wave energy $\left(\delta B^2/ B_0^2\right)$, and the relaxation of the plasma parameters: plasma $\left(\beta_{b,\parallel,\perp} \text{ and }\beta_{c,\parallel,\perp}\right)$,  and drift velocities $\left(u_{b, c}\right)$ of the beam and core components.}\label{f4}
\end{figure*}
%
\begin{figure*}
\centering
\includegraphics[scale=0.76, angle=-90, trim={6.5cm 4.cm 6.3cm 2.4cm}, clip]{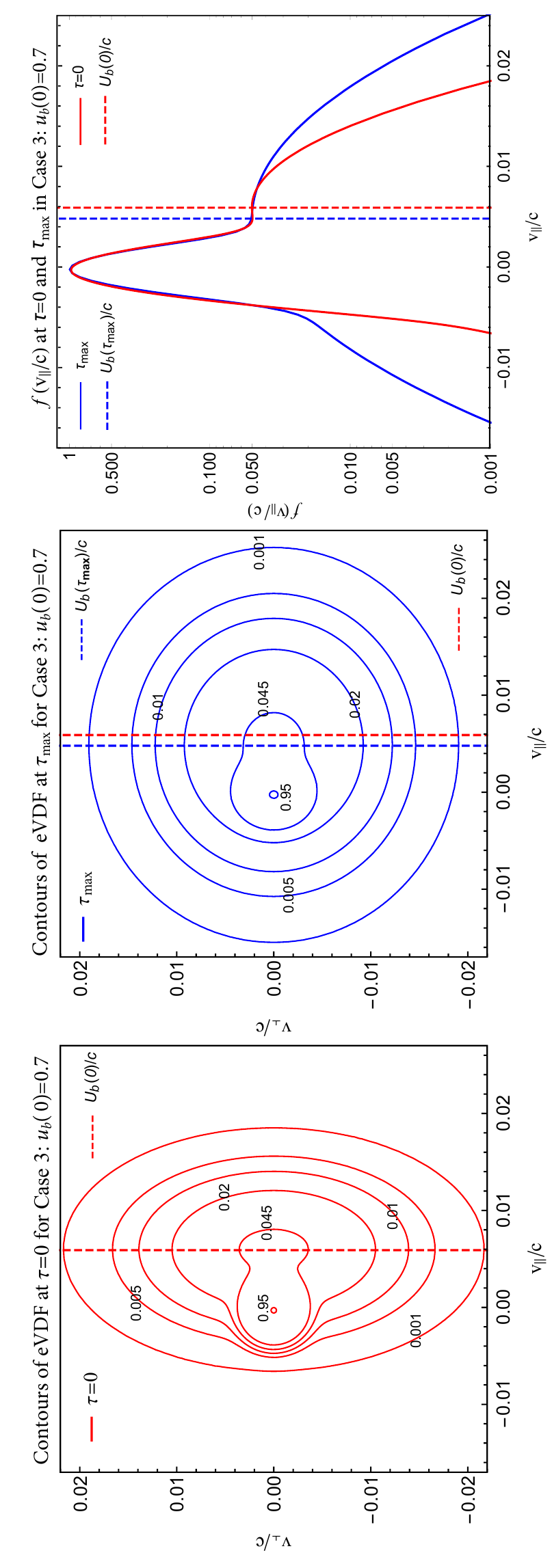}
\caption{Contours (left and middle panels) and parallel cuts (right) of eVDFs at $\tau=0$ (red) and $\tau_m$ (blue) from the same QL run for case~\ref{c3} with $u_b(0)=0.7$ in Fig.~\ref{f4}}\label{f5}
\end{figure*}
%
\subsubsection{Case~\ref{c2} -- anisotropic core electrons, $A_c(0)\neq 1$} 
%
In case \ref{c2}, we adopt the same initial plasma parameters as in case~\ref{c1}, but with anisotropic core populations 
$A_c(0)=A_b(0) \ne 1$. \cite{Shaaban2018a} have shown that growth rates of the WHF instability are only slightly 
changed under the influence of the core anisotropies, e.g., $A_c(0)=0.7-1.6$, see Fig~1 therein. Here in Figure~\ref{f3} 
we display temporal profiles of the wave energy density and the electron plasma parameters, which remain the same as in 
case \ref{c1}, excepting the core plasma beta $\beta_{\parallel,\perp c}$. In all these cases, the core population gains 
energy (heating) in perpendicular direction, as shown by the increase of the perpendicular component of plasma beta 
parameter (blue), but loses energy (cooling) in parallel direction (red), see Figure~\ref{f3}.  

\subsection{Whistler temperature anisotropy-driven instability}\label{sec.3.2}
In this section we consider different conditions of whistler instability (WI), as driven by sufficiently large temperature 
anisotropies $A_{b}(0)\equiv~\beta_{b~\perp}(0)/\beta_{b~\parallel}(0)=3.0$. In addition, an initial finite drift is assumed
between core and beam that may induce important effects on the enhanced wave fluctuations, inclusive on their saturation 
and macroscopic plasma parameters. By contrast to recent studies of WI \citep{Shaaban2019WI}, which 
consider only small beaming velocities, i.e.\ $u_b \leqslant 0.5$, here we assume $u_b(0)\geqslant~0.55$, for which the WHF mechanism
is expected to contribute to the instability leading to enhanced electromagnetic fluctuations.
First, in case~\ref{c3}, we minimize the effects of the core population, considering it isotropic, i.e., $A_c(0)=1.0$, in order 
to isolate the drift effects on the WI driven by the temperature anisotropy of beaming population. In the second part, 
in case~\ref{c4}, we examine the effects introduced by the core anisotropy $A_c(0)>1.0$.   

%
\subsubsection{Case~\ref{c3} -- counter-beaming electrons, $u_{c,b}(0)\neq 0$}
%
Fig.~\ref{f4} presents temporal evolution of the initial plasma parameters in case~\ref{c3}, for three different initial beaming 
velocities $u_b(0)=0.0$ (left), $0.55$ (middle), $0.7$ (right). For reference to previous studies, in the left panels we consider 
the case of an initially non-drifting electron populations with $u_{c,b}(0)=0.0$. The development of WI can regulate the initial 
temperature anisotropy of the beaming electrons, through the heating and cooling processes reflected by parallel (red) and 
perpendicular (blue) components of the electron plasma beta ($\beta_{\parallel,\perp,b}$). After the saturation of WI, i.e.\ at 
$\tau_m$, the beam electrons are less anisotropic, i.e., $A_b(\tau_m)\gtrsim1$, and may be at the origin of a WHF effect 
\citep{Shaaban2018a}. Initially isotropic ($A_c(0)=1.0$) the core electrons are subjected to parallel cooling (red) and 
perpendicular heating (blue), as shown by the core plasma beta parameters ($\beta_{\parallel, \perp, c}$), ending up with a 
small anisotropy induced in perpendicular direction, i.e.\ $A_c(\tau_m) \gtrsim 1.0$. For sufficiently large initial drifts 
$u_b(0)\geqslant~0.55$, temporal profiles of the macroscopic plasma parameters for both the beam and core populations, 
as well as the associated magnetic wave energy ($\delta B^2/B_0^2$) are markedly changed. An increase of the initial beaming 
velocities $u_b(0)\neq 0.0$, stimulates the heating and cooling processes for both the beam and core populations leading to 
higher levels of magnetic wave energy after the instability saturation. For instance, an initial beaming velocity $u_b(0)=0.55$, 
increases the relaxation of the beam anisotropy, making the beam electrons isotropic $A_b(t_{m})=1.0$ at later stages $\tau_{m}$. 
For initial beams with higher velocities, i.e., $u_b(0)=0.7$, their initial anisotropy $A_b(0)=3$  decreases passing through 
isotropic state $A_b(\tau)=1$ before reversing the anisotropy to $A_b(t_{m})<1.0$, later at $\tau_{m}$. The induced temperature 
anisotropy reached by the core electrons $A_c(\tau_m)>1.0$ increases for initial beams with higher velocities.  
It is important to mention that the relaxation of the anisotropic distributions, e.g., in Figures~\ref{f1}~and~\ref{f4}, 
is a complex process involving a reduction of the beaming velocity but at the same time a redistribution
of kinetic energy by thermalization, heating or cooling of electron populations. Instead, the magnetic  wave energy increases 
monotonously reaching a peak of saturation, and then decreases. This dynamical variation in the wave energy may be a common 
feature for the long-term evolution of a multi-component plasma, not only in theory \citep{Seough2012, Kim2017, Lazar2019} but 
also in hybrid and PIC simulations (e.g., \cite{Shoji2009, Bortnik2011, Kim2017, Lazar2019}), when the amplitude of the developed 
fluctuations is small enough to prevent nonlinear effects but still allow for differential interactions with electron populations, 
leading in this case not only to a  relaxation of beaming velocity, but also to core heating and beam thermalization. In numerical 
simulations the physical interpretation of this behavior is often attributed to the re-absorption of wave energy  after saturation, 
a phase which remains weakly nonlinear or even quasilinear (QL) if well reproduced by the QL approaches (see also the explanations 
in  \cite{Seough2012}, based on a competition between the damped  and amplifying ranges within the integration over wave-number $k$, 
like in our Eq.~(\ref{e6})). 

Based on the linear theory predictions both WHFI and WI are expected to develop with the same dispersive characteristics if
the initial beaming velocity is sufficiently large $u_b>0.5$, and the beam temperature anisotropy is $A_b(0)>1.2$ 
\citep{Shaaban2018a}. However, linear theory cannot identify the active regimes of these instabilities from their long-term
interplay. It is a QL analysis that may provide evidences for each of these WI or WHFI regime, and enables us to understand the 
interplay between different degrees or sources of free energy in the plasma system. For non-drifting electron populations 
$u_{b,c}(0)=0.0$ (serving as a reference) the temperature anisotropy of beaming electrons $A_b(0)\equiv \beta_{\perp,b}/
\beta_{\parallel,b}=3.0$ is partially relaxed by the enhanced fluctuations associated with the pure WI saturation, and this 
population ends up with lower but finite temperature anisotropy $A_b(\tau_m)=1.53$ at later stages, see left-top 
panel of Figure~\ref{f4}. Initial beaming velocities $u_b(0)\neq 0.0$ decelerate the relaxation process by a factor of $\sim3$ 
for $u_b=0.55$ and $\sim 6$ for $u_b=0.7$ compared to nondrifting case $u_b=0.0$, see gray lines in the middle- and right-top panels. This 
delay confirms the linear theory predictions for the inhibiting effect of counter-beams on the WI (decreasing the 
growth rates of WI by increasing the beaming velocity), see fig. 3 in \cite{Shaaban2018a}. In other words, for $u_b=0.55$ 
and $u_b=0.7$  the beam electrons need longer time to reach the same final anisotropy as for $u_b=0.0$. WHFI becomes 
operative and the relaxation of the beam anisotropy, as well as the wave energy density, both resemble the time evolution of a 
pure WHF in Figure~\ref{f1}~(right panels). As an evidence for the WIFI active regime, we observe the relaxation 
of the drift velocities for both the beam and core electrons only beyond the gray lines indicating the separation of 
two distinct regimes.

Figure~\ref{f5} displays contours of the eVDFs (left and middle panels) and their parallel cuts $f_e(v_\parallel/c)$ 
(right)  at initial $\tau=0$ (red) and final $\tau_m$ (blue) stages, corresponding to the results from the QL run in 
case~\ref{c3} with $u_b(0)=0.7$, see right panels in Figure~\ref{f4}. It is clear that after the instability saturation 
the eVDF is markedly different compared to the initial state, the beam electrons exhibit a small anisotropy in parallel 
direction and display a lower drift velocity. As a result, the eVDF at $\tau_m$ is less asymmetric and more stable against 
whistler instabilities.  

\begin{figure*}
\includegraphics[scale=1.1, trim={2.75cm 12.2cm 2.5cm 2.55cm}, clip]{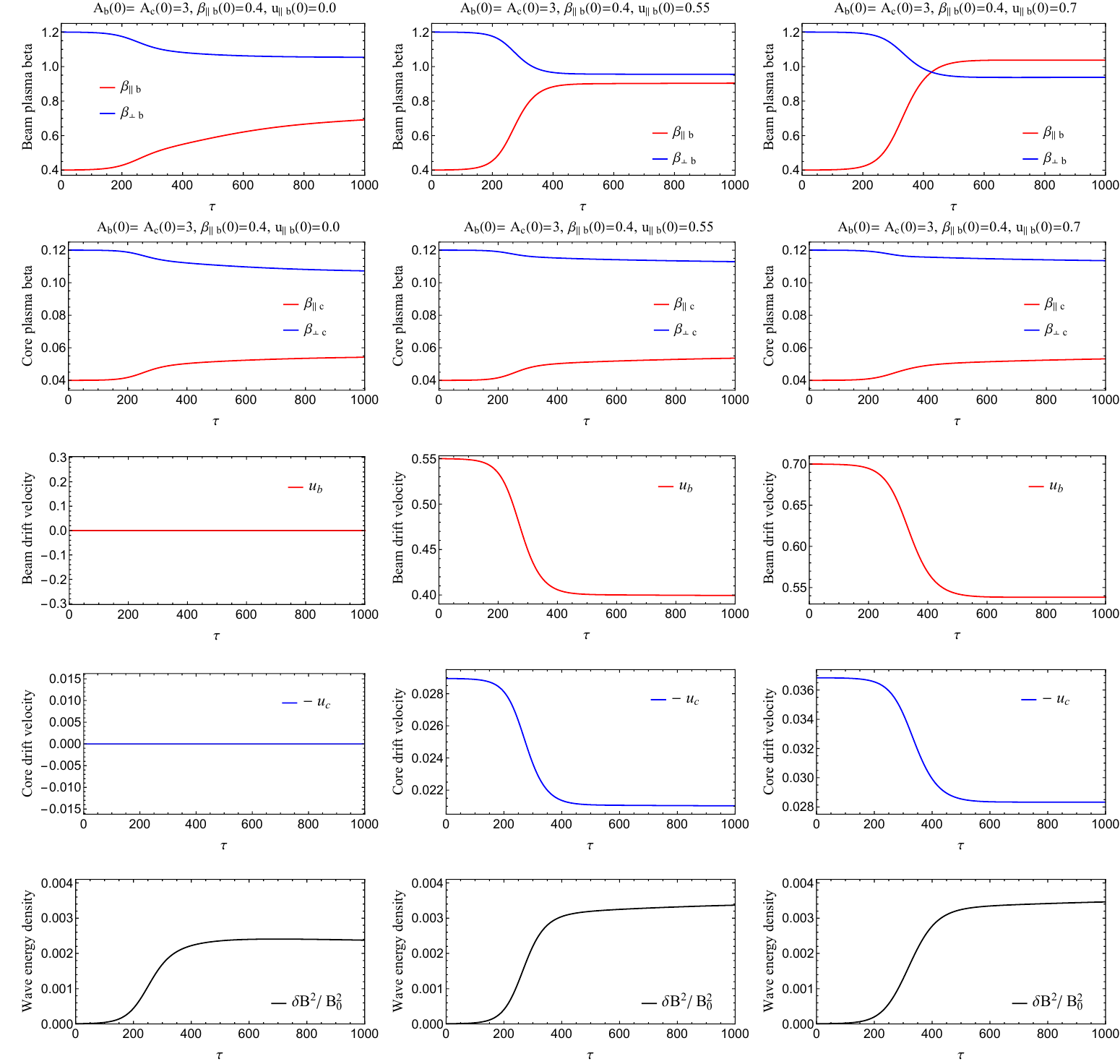}
\caption{Case~\ref{c4}: The same as in Figure~\ref{f3}, but with initially anisotropic core $A_c(0)=A_b(0)=3.0$.}\label{f6}
\end{figure*}
%
\subsubsection{Case~\ref{c4} -- drifting anisotropic core and beam electrons} 
%
Figure~\ref{f6} describes temporal evolution of the same initial plasma parameters assumed in case~\ref{c3}, but under 
the influence of an initially anisotropic core with $A_c(0)=A_b(0)=3$, which we name case~\ref{c4}. Again, we 
observe a reduction of temperature anisotropies, as well as drift velocities, but within a faster evolution
explained by the higher levels of wave energy density ($\delta B/B_0^2$). Initially anisotropic, the core electrons 
have direct consequences on how the drift velocity affects the temporal profiles of the plasma parameters, contrasting, 
for instance, with the results in Figure~\ref{f5}. The core electrons, initially with $A_c(0)=3.0$, determine a lower 
effectiveness of the drift velocities on the resulting instability, e.g., in the relaxation of temperature anisotropy, 
but  may boost the relaxation of the beam and core drift velocities. These results are consistent with predictions 
from linear theory that shows similar cumulative effects, i.e., markedly  enhanced growth-rates, from the interplay 
of the core and beam anisotropies and their relative drifts \citep{Lazar2018a, Shaaban2019WI}. 
The same growth rates of the WI are stimulated by the drift velocities, while the WHFI is only slightly enhanced
by the core anisotropy \citep{Shaaban2018a}.   
%

\begin{figure}
\centering
\includegraphics[scale=0.68, trim={4.5cm 4.8cm 2.5cm 3.2cm}, clip]{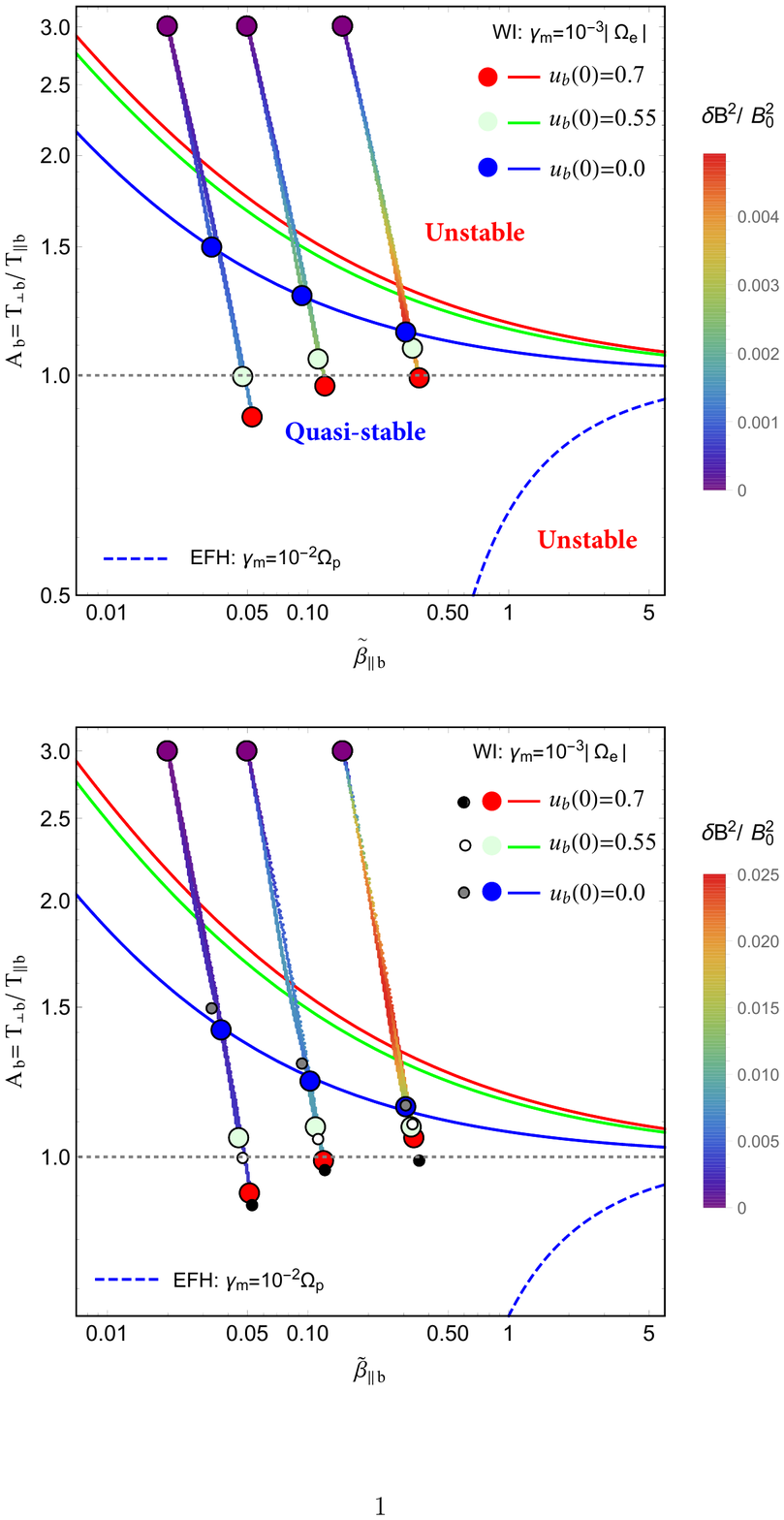}
\caption{Dynamical paths in parametric space ($A_b, \tilde{\beta}_{\parallel ~b}$) for beaming electrons 
for two initial conditions, $A_c(0)=1.0$ (top) and $A_c(0)=A_b(0)$ (bottom). Initial conditions are shown 
with purple filled circles, while final states are indicated with blue, green, and red filled circles. 
The magnetic wave energy density is color-coded. Contours corresponding to the anisotropy thresholds of
WI are plotted for different beaming velocities $u_b = 0.7$ (red), 0.55 (green), 0.0 (blue). Serving as 
a visual guidance, the threshold condition for the EFH instability (blue-dashed line) is taken from 
\citet{Shaaban2019EFH} for non-drifting electron populations.}\label{f7}
\end{figure}
%
\subsubsection{Electron temperature isotropization in the solar wind}\label{sec.3.2.3}
%
Fig.~\ref{f7} displays contours of temperature anisotropy ($A_j$) thresholds derived for finite maximum growth rates
$\gamma_m/|\Omega_e|=10^{-3}$, close to the marginal condition of stability (i.e.\ $\gamma \rightarrow 0$), and in 
terms of parallel plasma beta $\tilde{\beta}_{\parallel~j}\equiv~n_j~\beta_{\parallel~j}/n_0$. In order to
confirm the constraining effect on plasma particles these thresholds are usually compared with the limits of 
temperature anisotropy measured in space plasmas at time scales larger than those characteristic to kinetic 
instabilities \citep{Stverak2008, Lazar2017, Shaaban2019WI, Shaaban2019EFH, Bercic2019}. Here we refer only to 
beaming electrons (i.e., $j = b$), which are less dense but hotter than the core, and therefore more susceptible 
to deviations from isotropy. Contours of WI thresholds are fitted to an inverse power-law \citep{Lazar2015, Shaaban2019WI}
\begin{align}
A_{b}=1+\frac{a}{\tilde{\beta}_{\parallel b}^{~d}}\left(1-\frac{r}{\tilde{\beta}_{\parallel b}^ {~s}}\right),\label{th}
\end{align}
conditioning the anisotropy of the beam population ($A_b$) of the corresponding plasma beta ($\tilde{\beta}_{\parallel b} 
\equiv~n_b~\beta_{\parallel b}/n_0$). Fitting parameters $a$, $d$, $r$, and $s$ tabulated in Table~\ref{t1}.
These thresholds offer a plus of information regarding the variation of the instability growth-rates with different 
plasma parameters, in this case not only as a function of $\tilde{\beta}_{\parallel b}$, but also as a function of 
the core parameters, and the relative beaming velocity ($u_b$). Parametrization used in the present analysis is inspired from the observations
in the solar wind \citep{Maksimovic2005, Stverak2008, Pulupa2014, Tong2019a,Bercic2019}, where slow winds are
usually associated with lower beaming velocities, e.g., $u_b \simeq 0.0$, and more energetic events like fast winds
or coronal mass ejections may be characterized by higher beaming velocities $u_b=$0.55, 0.70. In the present notation,
beaming velocity $u_b=0.7$ implies $U_b\approx 30~v_A$ and $U_c\approx1.6 ~v_A$. Two panels in Figure~\ref{f7} 
correspond to different initial conditions, respectively, to $A_c(0)=1.0$ (top) and $A_c(0)=A_b(0)=3.0$ (bottom). 
The anisotropy thresholds of WI show a uniform variation increasing with $u_b$ and confirming the 
inhibiting effects of the relative drift \citep{Shaaban2018a}. For anisotropies $A_b < 1$, in the bottom-right 
corner, we show the threshold ($\gamma_m/\Omega_p=10^{-2}$) of the parallel firehose instability \citep{Shaaban2019EFH}, as 
an indication for the lower limits of anisotropy expected from the observations. 

However, the instability thresholds predicted by linear theory may have a reduced relevance for instabilities
resulted from the interplay of two plasma components \citep{Shaaban2019HF, Lopez2019}, as also 
in our present case. Such that, for the sake of completeness, in Figure~\ref{f7} we have added the results from our 
QL analysis, by considering 18 different combinations of initial conditions, implying different temperature 
anisotropies and parallel plasma betas. The initial conditions in cases \ref{c3} and \ref{c4} are multiplied 
by three different initial plasma betas $\beta_{\parallel b}(0)=0.4, 1.0$ and $3.0$ (implying 
$\tilde{\beta}_{\parallel~b}=0.02, 0.05, 0.15$). QL evolutions are shown as dynamical paths starting from initial 
positions, marked with purple filled circles, and ending after the instability saturation at final positions,
shown with blue, green, and red filled circles corresponding to the initial drift velocities $u_b(0)=0.0,~0.55$, 
and $0.7$. The wave energy density ($\delta B^2/B_0^2$) is coded with a \textit{rainbow} color scheme. In the 
top panel the core electrons are considered initially isotropic, in order to isolate the effects of drift 
velocities on the dynamical paths of temperature anisotropy. For $u_b=0.0$ final positions align perfectly to 
the corresponding anisotropy threshold predicted for WI (blue solid line) by the linear theory. For the other 
cases of finite $u_b(0)=0.55$ and $0.7$, the initially counter-streaming electron components may dramatically 
change the dynamical paths, which become longer to markedly lower temperature anisotropies approaching 
the isotropy $A_b =1$, or even below, changing to the opposite regime of $A_b \lesssim 1$. With increasing 
parallel plasma beta the dynamical paths become longer for $u=0.0$, while shorter for higher $u_b=0.55$ or $0.7$. 

Bottom panel in Figure~\ref{f7} show the case of an initially anisotropic core with $A_c(0)=3$. For the sake of 
comparison, with smaller circles filled in gray, white, and black we also show the final states obtained in top 
panel. Again, dynamical paths for $u_b=0.0$ end up exactly on the corresponding WI threshold predicted by linear 
theory, but for $u_b\neq 0.0$ the final states may slightly change, remaining however much below the corresponding 
linear thresholds and approaching a quasi-stable regime closer to $A_b \sim 1.0$. To be more exact, we can specify that for 
$u_b\neq~0.0$, the temperature anisotropy of the core population, $A_c(0)=A_b(0)=3.0$, determines slightly shorter 
dynamical paths, in agreement with the results in Figure~\ref{f6}. One possible explanation already mentioned in 
the previous sections is given by the interplay of different sources of free energies, which may trigger 
different instability mechanisms, i.e., WI or/and WHF, of the same branch of whistler modes (same dispersive characteristics).
Their cumulative effects may lead to enhanced fluctuations and more efficient effects of these fluctuations on
the relaxation of anisotropic electrons. 

These results may therefore offer a valuable alternative explanation for the observations of electron populations 
in the solar wind, which show the most stable states accumulating (highest number of events) nearly isotropic 
temperatures ($A \sim 1$) \citep{Stverak2008, Lazar2017}. The relative drift between different electron 
populations may exist even in the slow winds, and may therefore play a certain role in this case. Even the 
high-density core in the central population (subscript $j=c$) may respond to kinetic instabilities, 
which constrain their large deviations from isotropy \citep{Stverak2008, Lazar2017}. In this case the 
accumulation of quasi-stable states close to isotropy and equipartition of energy is usually attributed to another 
cumulative effect of binary collisions at lower heliospheric distances (e.g., in the outer corona), quantified by the so-called 
collisional age of plasma particles \citep{Salem2003, Stverak2008}. However, the electrons are much lighter and much 
faster than protons, and an explanation of their dynamics may not directly rely on the expansion and transport of the 
solar wind. 

\section{Conclusions}\label{Sec.4}

Counter-beaming populations of electrons are ubiquitous in the solar wind, and their interplay with the temperature anisotropies 
cannot be ignored \citep{Stverak2008, Vinas2010, Tong2019a}. Recent studies show that linear properties of whistler instabilities (i.e., 
WI and WHF, see above) are markedly affected the interplay of these sources of free energy \citep{Shaaban2018a, Tong2019a}. However, 
a linear approach is limited and cannot distinguish between different mechanisms responsible for instability or stabilization of 
enhanced fluctuations of the same plasma modes. We have carried out an advanced QL study of the whistler instabilities based on a 
parametric numerical analysis describing such complex (but less idealized) conditions resembling solar wind electron observations. 
Contrasting to previous studies, which restricted only to simplified approaches, e.g., 
(counter-)beaming Maxwellian populations with isotropic temperatures \citep{Shaaban2019HF, Lopez2019} or non-drifting 
populations with anisotropic temperatures \citep{Sarfraz2016, Lazar2018a}, our results describe 
the long-term evolution of whistler instability under the cumulative effects of both these sources of free energy.
In this case we can talk about more than two distinct regimes of the unstable fluctuations, the
one controlled mostly by the electron strahl (or electron heat flux) beam, the complementary regime dominated 
by the temperature anisotropy of beaming electrons, and an intermediary regime. A cumulative or intermediary regime can be 
identified when destabilizing effects of electron beam and of (small) temperature anisotropies are comparable, 
leading to highly stimulated fluctuations, and implicitly to a more efficient relaxation of the beaming electrons. 
Thus, our present QL analysis may provide a more realistic perspective enabling to understand more complex 
mechanisms responsible for the whistler fluctuations and their implications in the observations, see for instance 
\cite{Tong2019a, Tong2019b}.

In section \ref{sec.3.1} we have adopted two sets of initial plasma parameters, i.e.\ cases \ref{c1} and \ref{c2}, 
both favorable to the WHF instability. For an electron beam hotter but less dense than the core and 
with drift velocity lower than thermal speed the WHF instability can be self-generated \citep{Gary1975, Gary1985, 
Shaaban2018a}. The QL temporal evolution of the wave energy density ($\delta B^2/B_0^2$) and plasma parameters 
(core and beam drift velocities $u_{c,b}$, and beta parameters $\beta_{\perp j}$, $\beta_{\parallel j}$) are 
highly conditioned by the initial anisotropy of the beam population (Figure~\ref{f1}). A small (initial) 
anisotropy in perpendicular direction, i.e. $A_b(0)=1.2$, stimulates the instability and the resulting wave 
energy density, which determines a faster and deeper relaxation of the drift velocity, and induces a higher 
anisotropy to the core ($A_c > 1$). Parallel thermalization of the electron beam becomes also more pronounced 
leading to a complete relaxation of temperature anisotropy ($A_b = 1$), and later to a flip to opposite anisotropies 
($A_b <1$ or $\beta_{\perp, b}<\beta_{\parallel, b}$). If present, the initial anisotropy of the core has only a 
minor influence on the WHF fluctuations, including their saturation and the variation of plasma parameters (Figure~\ref{f3}). 

Section \ref{sec.3.2} investigates plasma conditions more favorable to WI, mainly driven by the temperature 
anisotropy, e.g., cases \ref{c3} and \ref{c4}. Initially isotropic ($A_c(0)=1.0$, case~\ref{c3}), the core 
electrons gain energy from the enhanced WI fluctuations, and become anisotropic in perpendicular direction 
at later stages, i.e., $A_c(\tau_m)>1.0$. An initial relative drift $u_b \neq 0$ is another key factor that 
may stimulate the enhanced fluctuations and implicitly the relaxation of the anisotropy through 
pitch-angle scattering leading to effective perpendicular cooling and parallel heating of beaming electrons. 
The relaxation of temperature anisotropy becomes more pronounced in this case (Figure~\ref{f4}), and 
for more energetic beams ($u_b = 0.7$) may end up with an opposite anisotropy after saturation. 
Temporal evolutions of the enhanced fluctuations and the macroscopic plasma parameters suggest transitions 
from one regime to another. For $u_b(0)\geqslant 0.55$, at early stages dominant is WI and the beam anisotropy 
is partially relaxed ($A_b(0)>A_b(\tau)>1$) by the enhanced WI fluctuations, while at later stages WI is 
saturated and apparently the WHF instability becomes operative, leading to a complete relaxation of the 
temperature anisotropy ($A_b(\tau_m)\lesssim 1$). An initially anisotropic core $A_c(0)=A_b(0)=3.0$ 
(case~\ref{c4}) reduces the instability effect of beaming electrons, but may increase the level of the 
fluctuating magnetic field, confirming a cumulative WI of the core and beam anisotropies predicted by linear theory. 

As another related application of our QL approach, in section~\ref{sec.3.2.3} we have considered the problem of 
temperature isotropization in the solar wind. In the observations of electron populations the highest number of events 
accumulate at the quasi-stable states nearly isotropic temperatures ($A \sim 1$), and below the instability thresholds 
predicted by linear theory \citep{Stverak2008, Lazar2017}. Figures~\ref{f7} compares these thresholds derived 
from linear theory and dynamical paths of the beam anisotropy from our QL computations. For non-drifting 
electron populations $u_j(0)=0$ the final positions of the dynamical paths settle down, exactly on the corresponding 
anisotropy thresholds, while for $u_j(0) \neq 0$ the dynamical paths extend below the linear anisotropy thresholds 
to very small anisotropies which approach the isotropy conditions $A_b\lesssim 1$. To conclude, our present 
findings suggest that whistler instabilities cumulatively driven by multiple sources of free energy are
expected to contribute to a more efficient relaxation of the anisotropic electrons, and may therefore provide a 
valuable alternative explanation for the isotropization of the solar wind electrons.

%
\begin{table}
        \centering
        \caption{
         Fitting parameters in Eq.~(\ref{th}) for WI with $\gamma_m=10^{-3} |\Omega_e|$}
   \label{t1}
        \begin{tabular}{lcccccccc} 
                \hline
                &   \multicolumn{2}{c}{$u_b(0)=0.0$} & \multicolumn{2}{c}{$u_b(0)=0.55$} & \multicolumn{2}{c}{$u_b(0)=0.7$}\\      
               $A_c(0)$ & 3& 1 & 3 & 1 & 3 & 1 \\
                \hline
                $a$ & 0.07 & 0.078& 0.16 & 0.157  & 0.18 & 0.18 \\
                $d$ & 0.54 & 0.54  & 0.48 & 0.49  & 0.47 & 0.48\\
                $r/10^{-5}$ & 1& 2  &2 &  3& 1 &  2\\
                $s$ & 1.0    & 1.0    & 1.0 & 1.0  & 1.0 & 1.0\\
                \hline
        \end{tabular}
\end{table}

\section*{Acknowledgements}
These results were obtained in the framework of the projects SCHL 201/35-1 (DFG-German Research Foundation), 
GOA/ 2015-014 (KU Leuven), G0A2316N (FWO-Vlaanderen). S.M. Shaaban acknowledges support by a FWO Postdoctoral 
Fellowship (Grant No. 12Z6218N). Thanks are also due to Rodrigo~A.~L{\'o}pez and Peter~H.~Yoon for insightful discussions in the framework
of our joint research projects (Grant No. SF/17/007-2018, ISSI project on Kappa Distributions 2017-2019). 




\bibliographystyle{mnras}
\bibliography{papers2} 


\bsp	
\label{lastpage}
\end{document}